\documentclass[epj]{svjour}
\usepackage{graphics}
\def\graph#1{
\begin{array}{c}\mbox{
\resizebox{0.06\textheight}{!}{%
\includegraphics*{#1.eps}
                               }
}\end{array}
            }
\def\graphL#1{
\begin{array}{c}\mbox{
\resizebox{0.1\textheight}{!}{%
\includegraphics*{#1.eps}
                               }
}\end{array}
             }
\def\graphS#1{
\begin{array}{c}\mbox{
\resizebox{0.03\textheight}{!}{%
\includegraphics*{#1.eps}
                               }
}\end{array}
             }
\begin{document}

\newcommand {\nn}{\nonumber \\}
\newcommand {\Tr}{{\rm Tr\,}}
\newcommand {\tr}{{\rm tr\,}}
\newcommand {\e}{{\rm e}}
\newcommand {\etal}{{\it et al.}}
\newcommand {\m}{\mu}
\newcommand {\n}{\nu}
\newcommand {\pl}{\partial}
\newcommand {\p} {\phi}
\newcommand {\vp}{\varphi}
\newcommand {\vpc}{\varphi_c}
\newcommand {\al}{\alpha}
\newcommand {\be}{\beta}
\newcommand {\ga}{\gamma}
\newcommand {\Ga}{\Gamma}
\newcommand {\x}{\xi}
\newcommand {\ka}{\kappa}
\newcommand {\la}{\lambda}
\newcommand {\La}{\Lambda}
\newcommand {\si}{\sigma}
\newcommand {\Si}{\Sigma}
\newcommand {\sh}{\theta}
\newcommand {\Th}{\Theta}
\newcommand {\om}{\omega}
\newcommand {\Om}{\Omega}
\newcommand {\ep}{\epsilon}
\newcommand {\vep}{\varepsilon}
\newcommand {\na}{\nabla}
\newcommand {\del}  {\delta}
\newcommand {\Del}  {\Delta}
\newcommand {\mn}{{\mu\nu}}
\newcommand {\ls}   {{\lambda\sigma}}
\newcommand {\ab}   {{\alpha\beta}}
\newcommand {\gd}   {{\gamma\delta}}
\newcommand {\half}{ {\frac{1}{2}} }
\newcommand {\third}{ {\frac{1}{3}} }
\newcommand {\fourth} {\frac{1}{4} }
\newcommand {\sixth} {\frac{1}{6} }
\newcommand {\sqtwo} {\sqrt{2}}
\newcommand {\sqg} {\sqrt{g}}
\newcommand {\fg}  {\sqrt[4]{g}}
\newcommand {\invfg}  {\frac{1}{\sqrt[4]{g}}}
\newcommand {\sqZ} {\sqrt{Z}}
\newcommand {\sqk} {\sqrt{\kappa}}
\newcommand {\sqt} {\sqrt{t}}
\newcommand {\sql} {\sqrt{l}}
\newcommand {\reg} {\frac{1}{\epsilon}}
\newcommand {\fpisq} {(4\pi)^2}
\newcommand {\Acal}{{\cal A}}
\newcommand {\Lcal}{{\cal L}}
\newcommand {\Ocal}{{\cal O}}
\newcommand {\Dcal}{{\cal D}}
\newcommand {\Ncal}{{\cal N}}
\newcommand {\Mcal}{{\cal M}}
\newcommand {\scal}{{\cal s}}
\newcommand {\Dvec}{{\hat D}}   
\newcommand {\dvec}{{\vec d}}
\newcommand {\Evec}{{\vec E}}
\newcommand {\Hvec}{{\vec H}}
\newcommand {\Vvec}{{\vec V}}
\newcommand {\lsim}{
\vbox{\baselineskip=4pt \lineskiplimit=0pt \kern0pt 
\hbox{$<$}\hbox{$\sim$}}
                    }
\newcommand {\rpl}{{\vec \partial}}
\def\overleftarrow#1{\vbox{\ialign{##\crcr
 $\leftarrow$\crcr\noalign{\kern-1pt\nointerlineskip}
 $\hfil\displaystyle{#1}\hfil$\crcr}}}
\def\lpl{{\overleftarrow\partial}}
\newcommand {\Btil}{{\tilde B}}
\newcommand {\atil}{{\tilde a}}
\newcommand {\btil}{{\tilde b}}
\newcommand {\ctil}{{\tilde c}}
\newcommand {\dtil}{{\tilde d}}
\newcommand {\Ftil}{{\tilde F}}
\newcommand {\Ktil}  {{\tilde K}}
\newcommand {\Ltil}  {{\tilde L}}
\newcommand {\mtil}{{\tilde m}}
\newcommand {\ttil} {{\tilde t}}
\newcommand {\Qtil}  {{\tilde Q}}
\newcommand {\Rtil}  {{\tilde R}}
\newcommand {\Stil}{{\tilde S}}
\newcommand {\Ztil}{{\tilde Z}}
\newcommand {\altil}{{\tilde \alpha}}
\newcommand {\betil}{{\tilde \beta}}
\newcommand {\etatil} {{\tilde \eta}}
\newcommand {\latil}{{\tilde \lambda}}
\newcommand {\Latil}{{\tilde \Lambda}}
\newcommand {\ptil}{{\tilde \phi}}
\newcommand {\Ptil}{{\tilde \Phi}}
\newcommand {\natil} {{\tilde \nabla}}
\newcommand {\xitil} {{\tilde \xi}}
\newcommand {\Hbtil} {{\Huge {\~{b}}}}
\newcommand {\Hctil} {{\Huge {\~{c}}}}
\newcommand {\Hdtil} {{\Huge {\~{d}}}}
\newcommand {\Ahat}{{\hat A}}
\newcommand {\ahat}{{\hat a}}
\newcommand {\Rhat}{{\hat R}}
\newcommand {\Shat}{{\hat S}}
\newcommand {\ehat}{{\hat e}}
\newcommand {\mhat}{{\hat m}}
\newcommand {\shat}{{\hat s}}
\newcommand {\Dhat}{{\hat D}}   
\newcommand {\Vhat}{{\hat V}}   
\newcommand {\xhat}{{\hat x}}
\newcommand {\Zhat}{{\hat Z}}
\newcommand {\Gahat}{{\hat \Gamma}}
\newcommand {\Phihat} {{\hat \Phi}}
\newcommand {\phihat} {{\hat \phi}}
\newcommand {\vphat} {{\hat \varphi}}
\newcommand {\nah} {{\hat \nabla}}
\newcommand {\etahat} {{\hat \eta}}
\newcommand {\omhat} {{\hat \omega}}
\newcommand {\psihat} {{\hat \psi}}
\newcommand {\thhat} {{\hat \theta}}
\newcommand {\gh}  {{\hat g}}
\newcommand {\abar}{{\bar a}}
\newcommand {\Abar}{{\bar A}}
\newcommand {\cbar}{{\bar c}}
\newcommand {\bbar}{{\bar b}}
\newcommand {\gbar}{\bar{g}}
\newcommand {\Bbar}{{\bar B}}
\newcommand {\Dbar}{{\bar D}}
\newcommand {\fbar}{{\bar f}}
\newcommand {\Fbar}{{\bar F}}
\newcommand {\kbar}  {{\bar k}}
\newcommand {\Kbar}  {{\bar K}}
\newcommand {\Lbar}  {{\bar L}}
\newcommand {\Qbar}  {{\bar Q}}
\newcommand {\Wbar}  {{\bar W}}
\newcommand {\albar}{{\bar \alpha}}
\newcommand {\bebar}{{\bar \beta}}
\newcommand {\epbar}{{\bar \epsilon}}
\newcommand {\labar}{{\bar \lambda}}
\newcommand {\psibar}{{\bar \psi}}
\newcommand {\vpbar}{{\bar \varphi}}
\newcommand {\Psibar}{{\bar \Psi}}
\newcommand {\Phibar}{{\bar \Phi}}
\newcommand {\chibar}{{\bar \chi}}
\newcommand {\sibar}{{\bar \sigma}}
\newcommand {\xibar}{{\bar \xi}}
\newcommand {\thbar}{{\bar \theta}}
\newcommand {\Thbar}{{\bar \Theta}}
\newcommand {\bbartil}{{\tilde {\bar b}}}
\newcommand {\aldot}{{\dot{\alpha}}}
\newcommand {\bedot}{{\dot{\beta}}}
\newcommand {\gadot}{{\dot{\gamma}}}
\newcommand {\deldot}{{\dot{\delta}}}
\newcommand {\alp}{{\alpha'}}
\newcommand {\bep}{{\beta'}}
\newcommand {\gap}{{\gamma'}}
\newcommand {\bfZ} {{\bf Z}}
\newcommand {\BFd} {{\bf d}}
\newcommand  {\vz}{{v_0}}
\newcommand  {\ez}{{e_0}}
\newcommand  {\mz}{{m_0}}
\newcommand  {\xf}{{x^5}}
\newcommand  {\yf}{{y^5}}
\newcommand  {\Zt}{{Z$_2$}}
\newcommand {\intfx} {{\int d^4x}}
\newcommand {\intdX} {{\int d^5X}}
\newcommand {\inttx} {{\int d^2x}}
\newcommand {\change} {\leftrightarrow}
\newcommand {\ra} {\rightarrow}
\newcommand {\larrow} {\leftarrow}
\newcommand {\ul}   {\underline}
\newcommand {\pr}   {{\quad .}}
\newcommand {\com}  {{\quad ,}}
\newcommand {\q}    {\quad}
\newcommand {\qq}   {\quad\quad}
\newcommand {\qqq}   {\quad\quad\quad}
\newcommand {\qqqq}   {\quad\quad\quad\quad}
\newcommand {\qqqqq}   {\quad\quad\quad\quad\quad}
\newcommand {\qqqqqq}   {\quad\quad\quad\quad\quad\quad}
\newcommand {\qqqqqqq}   {\quad\quad\quad\quad\quad\quad\quad}
\newcommand {\lb}    {\linebreak}
\newcommand {\nl}    {\newline}

\newcommand {\vs}[1]  { \vspace*{#1 cm} }
\newcommand {\IJMP}  {Int.Jour.Mod.Phys.}


%
\title{Graphical Representation of SUSY and Application to QFT}
\author{Shoichi Ichinose
}                     
\offprints{}          
\institute{Laboratory of Physics,
School of Food and Nutritional Sciences,
University of Shizuoka,Yada 52-1,Shizuoka 422-8525,Japan; 
\email{ichinose@smail.u-shizuoka-ken.ac.jp}}
\date{Received: date / Revised version: date}
%
\abstract{
We present a graphical representation of the supersymmetry
and the graphical calculation. 
Calculation is demonstrated for 4D Wess-Zumino model and for
Super QED.
The chiral operators are graphically expressed
in an illuminating way. The tedious part of SUSY calculation, due to
manipulating chiral suffixes, reduces considerably. The application
is diverse.
\PACS{
      {02.10.Ox}{Copmbinatorics, graph theory}   \and
      {02.70.-c}{Computational techniques, ...}
     } 
} 
\maketitle
%
The supersymmetry is the symmetry between fermions and bosons. 
It was introduced in the mid 70's. At present the experiment
does not yet confirm the symmetry, but everybody accepts its importance
in nature and expects fruitful results in the future developement. 
The requirement of such a high symmetry costs a sophisticated
structure which makes its {\it dynamical} analysis difficult. 
In this circumstance, we propose a calculational technique which
utilizes the graphical representation of SUSY. 
The representation was proposed in \cite{SI03,SUSY2004}.
\footnote{
An improved version of Ref.\cite{SI03} has recently appeared as Ref.\cite{SI06UW07}.
Details of the present article are given in Ref.\cite{SI06UW08}. 
} 
The spinor is represented as a slanted line with a direction. Its chirality
is represented by the way the line is drawn. The advantage of the graph expression
is the use of the {\it graph indices}. Every independent graph, which corresponds
to a unique term in the ordinary calculation, is classified by a set of graph indices.
Hence the main efforts of programinng is devoted to find good graph indices
and to count them. SUSY calculation
generally is not a simple algebraic or combinatoric or analytical one. 
It involves the vast branch of mathematics including Grassmann algebra.
The delicate property of chirality is produced in this environment. 
It requires a basic language for flexible programming.
We take C-language and present the output of a first-step program.

Weyl spinors have the SU(2)$_L \times$SU(2)$_R$ structure.
The {\it chiral} suffix $\al$, appearing in $\psi^\al$ or $\psi_\al$, 
represents (fundamental representation, doublet representation)
SU(2)$_L$ and the {\it anti-chiral} suffix $\aldot$, appearing in
$\psibar^\aldot$ or $\psibar_\aldot$, represents SU(2)$_R$. 
The raising and lowering of suffixes are done
by the antisymmetric tensors $\ep^{\al\be}$ and $\ep_{\al\be}$.
\begin{eqnarray}
(\ep^{\al\be})=
\left(
\begin{array}{cc}
0 & 1 \\
-1 & 0
\end{array}
\right),\ 
(\ep_{\aldot\bedot})=
\left(
\begin{array}{cc}
0 & -1 \\
1 & 0
\end{array}
\right),\ 
\ep^{\al\be}\ep_{\be\ga}=\del^\al_\ga ,\nn
\psi^\al=\ep^{\al\be}\psi_\be,\  
\psibar_\aldot=\ep_{\aldot\bedot}\psibar^\bedot
.
\label{int1}
\end{eqnarray}
They are graphically expressed by Fig.1. 
\begin{figure}
\resizebox{0.4\textwidth}{!}{%
\includegraphics{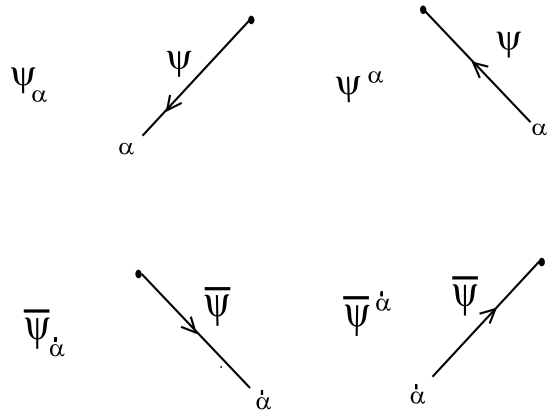}
                             }
\label{LF1}
   \caption{
Weyl fermions.
   }
\end{figure}
We encode them as follows. We use 2 dimensional array
with the size 2$\times$2. The four chiral spinors 
are stored in C-program as the array psi[\ ][\ ].

(1) (Weyl) Spinor [Symbol: p\ ;\ Dimension: M$^{3/2}$ ]\nl
\shortstack[l]{
$\psi^\al$\\
psi[0,0]=$\al$\\
psi[0,1]=emp\\
psi[1,0]=emp\\
psi[1,1]=emp
             }  
\q
\shortstack[l]{
$\psi_\al$\\
psi[0,0]=emp\\
psi[0,1]=$\al$\\
psi[1,0]=emp\\
psi[1,1]=emp
             }
\q
\shortstack[l]{
$\psibar_\aldot$\\
psi[0,0]=emp\\
psi[0,1]=emp\\
psi[1,0]=emp\\
psi[1,1]=$\aldot$
             }
\q
\shortstack[l]{
$\psibar^\aldot$\\
psi[0,0]=emp\\
psi[0,1]=emp\\
psi[1,0]=$\aldot$\\
psi[1,1]=emp
             }

The first column takes two numbers 0 and 1; 0 expresses a 'chiral' operator
$\psi$, while 1 expresses an 'anti-chiral' operator $\psibar$. The second
column also takes the two numbers; 0 expresses an 'up' suffix, while 1 expresses
an 'down' one. 

Note:\ 'emp' means 'empty' and is expressed by a default number (, for example, 99) in the program.  \nl
\nl

(2) Sigma Matrix [Symbol: s\ ;\ Dimension: M$^0$ ]\nl
\q Sigma matrices $\si^m, \sibar^m$ are graphically expressed in Fig.2.\nl 
\begin{figure}
\resizebox{0.3\textheight}{!}{%
\includegraphics{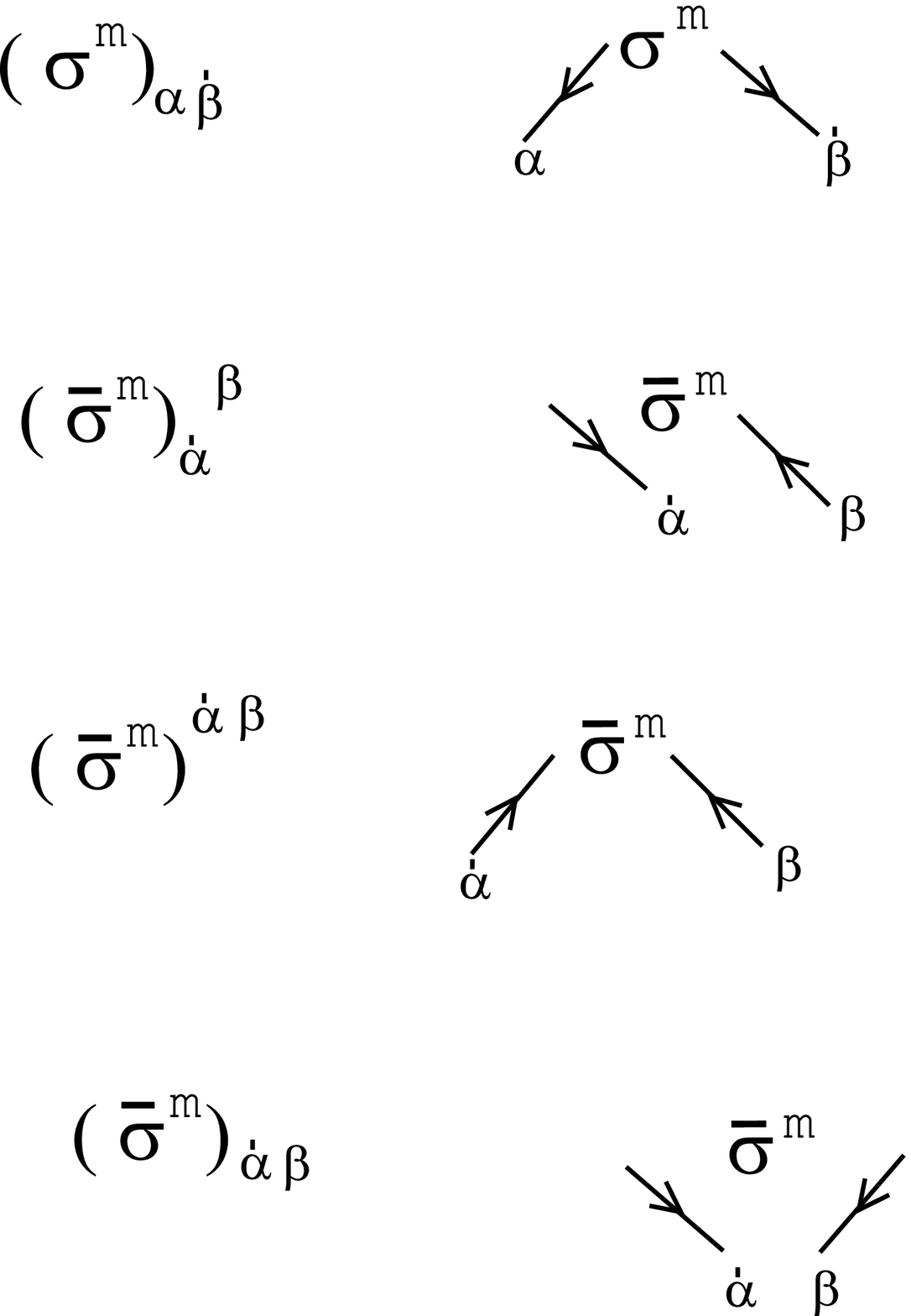}
                             }
\label{LF3}
   \caption{
Elements of SL(2,C) $\si$-matrices. 
$(\si^m)_{\al\bedot}$ and $(\sibar^m)^{\aldot\be}$ are
the standard form.
   }
\end{figure}
They are stored as the 2$\times$2 array si[\ ][\ ].\nl
\shortstack[l]{
$\si^m_{~\al\aldot}$\\
si[0,0]=emp\\
si[0,1]=$\al$\\
si[1,0]=emp\\
si[1,1]=$\aldot$\\
siv=m
             }
\q
\shortstack[l]{
$\sibar^{m\aldot\al}$\\
si[0,0]=$\aldot$\\
si[0,1]=emp\\
si[1,0]=$\al$\\
si[1,1]=emp\\
siv=m
             }
\q

(3)Superspace coordinate[Symbol: t;\ Dimension: M$^{-1/2}$ ]\nl
The superspace coordinate $\sh^\al$ is exprssed in the same way
as the spinor $\psi^\al$.\nl
\nl
\shortstack[l]{
$\sh^\al$\\
th[0,0]=$\al$\\
th[0,1]=emp\\
th[1,0]=emp\\
th[1,1]=emp
             }  
\q
\shortstack[l]{
$\sh_\al$\\
th[0,0]=emp\\
th[0,1]=$\al$\\
th[1,0]=emp\\
th[1,1]=emp
             }
\q
\shortstack[l]{
$\thbar_\aldot$\\
th[0,0]=emp\\
th[0,1]=emp\\
th[1,0]=emp\\
th[1,1]=$\aldot$
             }
\q
\shortstack[l]{
$\thbar^\aldot$\\
th[0,0]=emp\\
th[0,1]=emp\\
th[1,0]=$\aldot$\\
th[1,1]=emp
             }
\nl
They are graphically expressed by Fig.3.
\begin{figure}
\resizebox{0.3\textheight}{!}{%
\includegraphics{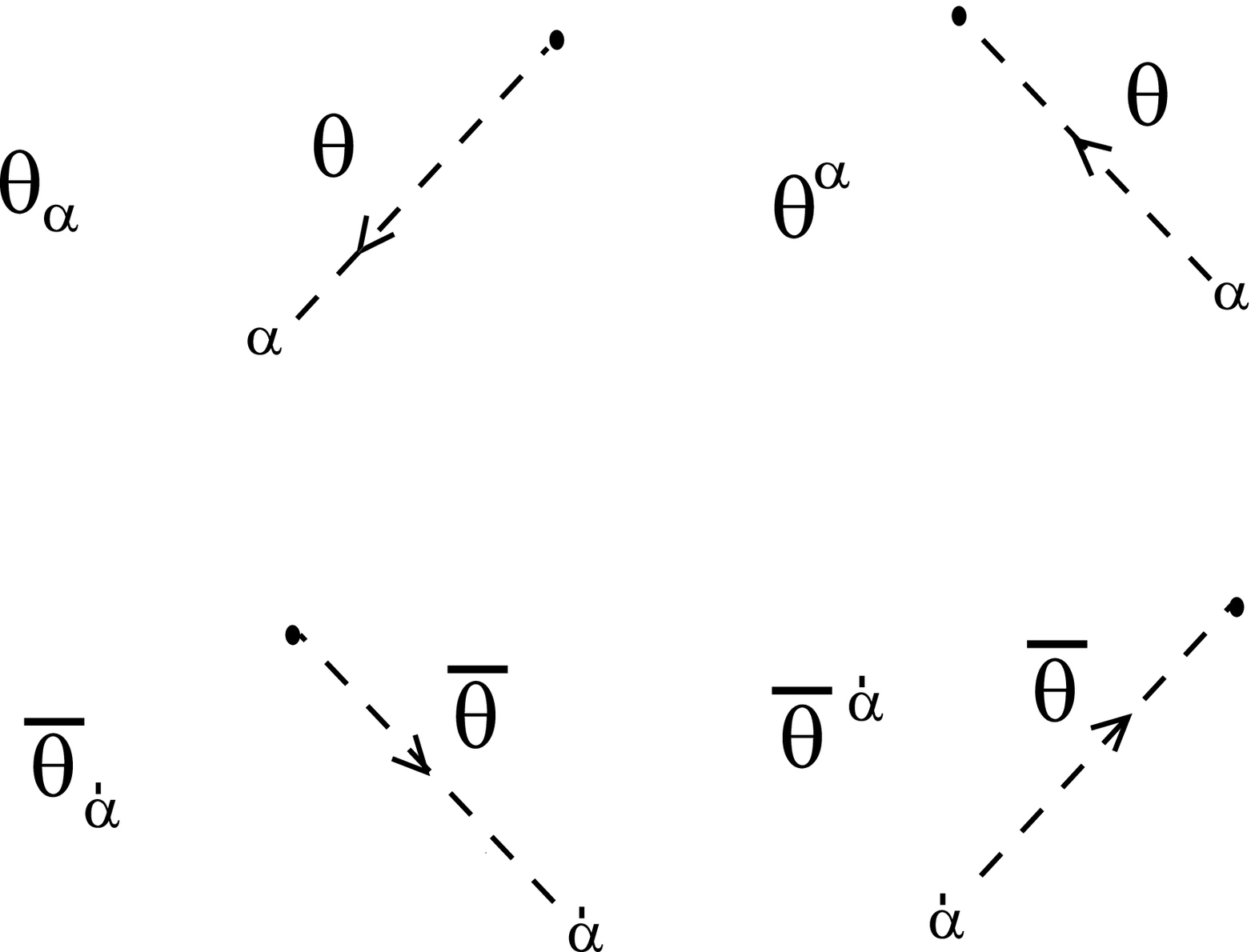}
                              }
\label{LF15}
   \caption{
The graphical representation for the
spinor coordinates in the superspace:\ $\sh_\al, \sh^\al, \thbar_\aldot$ and 
$\thbar^\aldot$.
   }
\end{figure}

(4) Gagino [Symbol: l\ ;\ Dimension: M$^{3/2}$ ]\nl
The photino $\la^\al$ is exprssed in the same way
as the spinor $\psi^\al$.
We take the 2$\times$2 array la[\ ][\ ].\nl
\nl
\shortstack[l]{
$\la^\al$\\
la[0,0]=$\al$\\
la[0,1]=emp\\
la[1,0]=emp\\
la[1,1]=emp
             }  
\q
\shortstack[l]{
$\la_\al$\\
la[0,0]=emp\\
la[0,1]=$\al$\\
la[1,0]=emp\\
la[1,1]=emp
             }
\q
\shortstack[l]{
$\labar_\aldot$\\
la[0,0]=emp\\
la[0,1]=emp\\
la[1,0]=emp\\
la[1,1]=$\aldot$
             }
\q
\shortstack[l]{
$\labar^\aldot$\\
la[0,0]=emp\\
la[0,1]=emp\\
la[1,0]=$\aldot$\\
la[1,1]=emp
             }


In the process of SUSY calculation, there appear graphs connected by
directed lines (chiral suffixes contraction) and by (non-directed) dotted lines
(vector suffixes contraction). We can classify them by some {\it graph indices}:\  
{\bf (1)vpairno} The number of  vector-suffix contractions;\ 
{\bf (2)NcpairO} The number of chiral-suffix contractions. This is equal to the number
of left-directed wedges;\ 
{\bf (3)NcpairE} The number of anti-chiral-suffix contractions. This is equal to the number
of the right-directed wedges;\ 
{\bf (4)closed-chiral-loop-No} The closed-chiral-loop is the case that the directed lines 
, connected by $\si$ or $\sibar$, make a loop. In this case NcpairO=NcpairE. 
The number of closed chiral loops is defined to this index;\  
{\bf (5)GrNum} A group is defined to be a set of $\si$'s or $\sibar$'s which
are connected by directed lines. The number of groups is defined to be GrNum.

In TABLE 1-2, we list the classification of the product of $\si$'s using
the graph indices defined above.
\begin{table}
\begin{center}
\begin{tabular}{c|c|c|c}
\hline
vpairno & NcpairO & NcpairE & figure\\
\hline
      &   0     &    0     &  $\graph{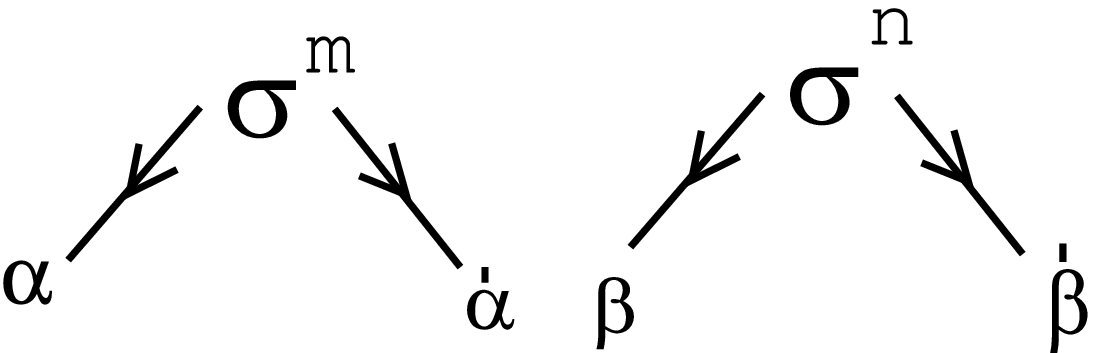}$  \\
\cline{2-4}
  0   &   0      &   1     &  $\graph{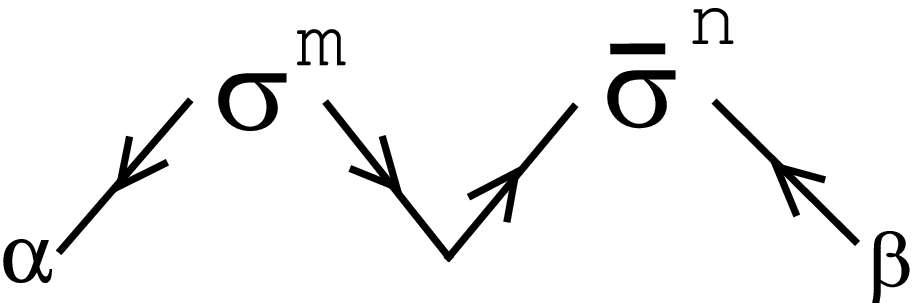}$  \\
\cline{2-4}
       &   1     &   0      &  $\graph{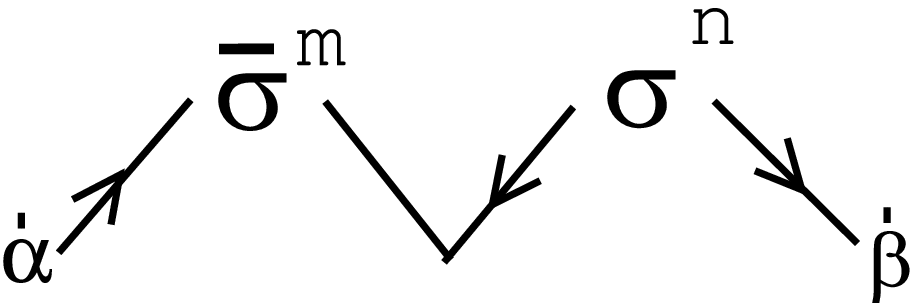}$  \\
\cline{2-4}
       &   1      &   1     &  $\graph{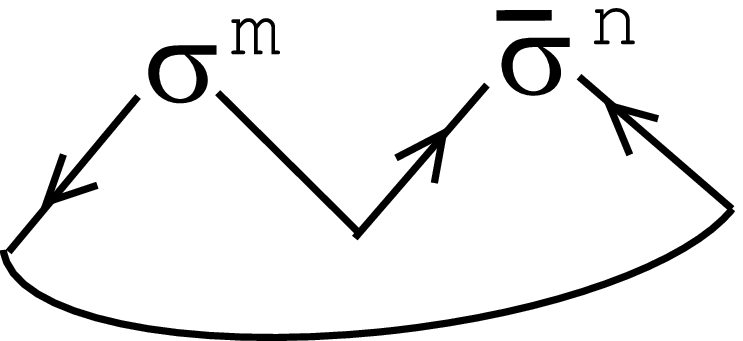}\ =-2\eta^{mn}$  \\
\hline
       &   0       &  0      &  $\graph{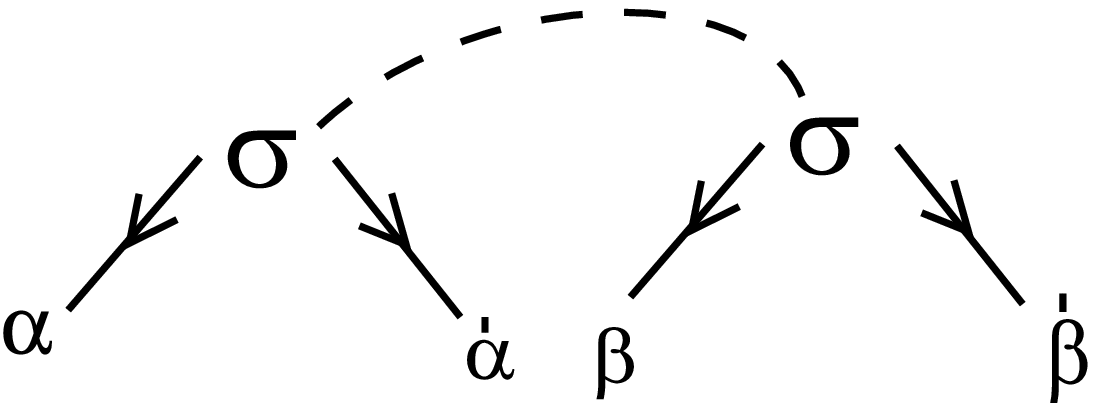}\ =-2\ep_{\al\be}\ep_{\aldot\bedot}$  \\
\cline{2-4}
   1   &   0      &    1     &  $\graph{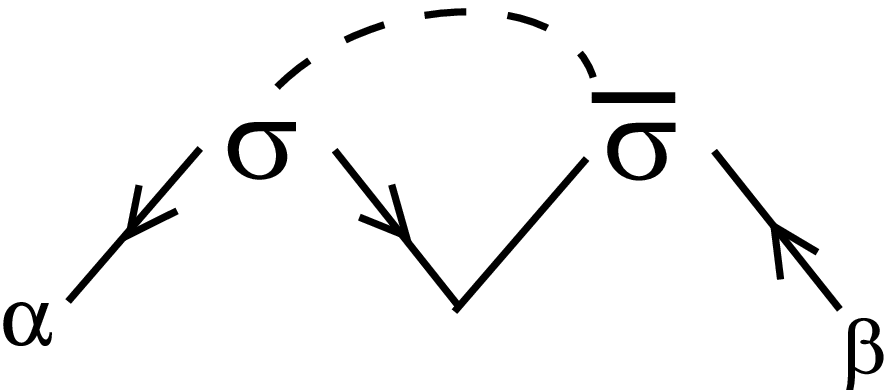}\ =-4\del^\be_\al$  \\
\cline{2-4}
       &    1      &    0     &  $\graph{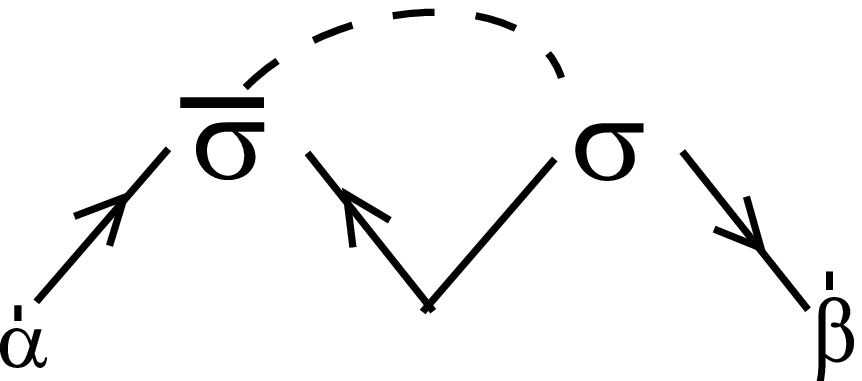}\ =-4\del_\bedot^\aldot$  \\
\cline{2-4}
        &    1     &    1     &   $\graph{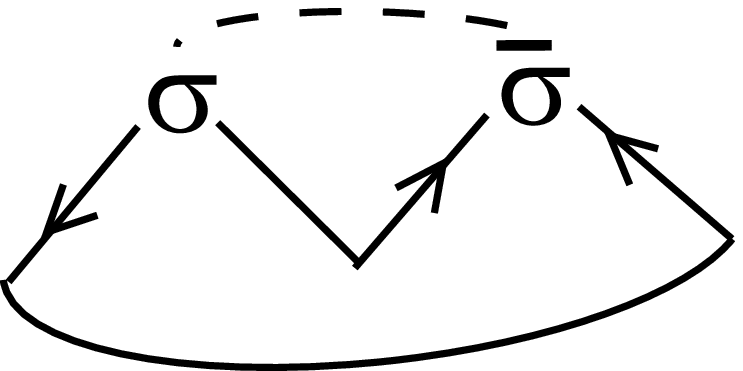}\ =-8$  \\
\hline
\multicolumn{4}{c}{}\\
\multicolumn{4}{c}{TABLE 1\ Classification of the product of 2 sigma matrices.
                  }
\end{tabular}
\end{center}
\end{table}
\begin{table}
\begin{center}
\begin{tabular}{c|c|c|c}
\hline
vpairno & NcpairO & NcpairE & figure\\
\hline
      &   0     &    0     &  $\graph{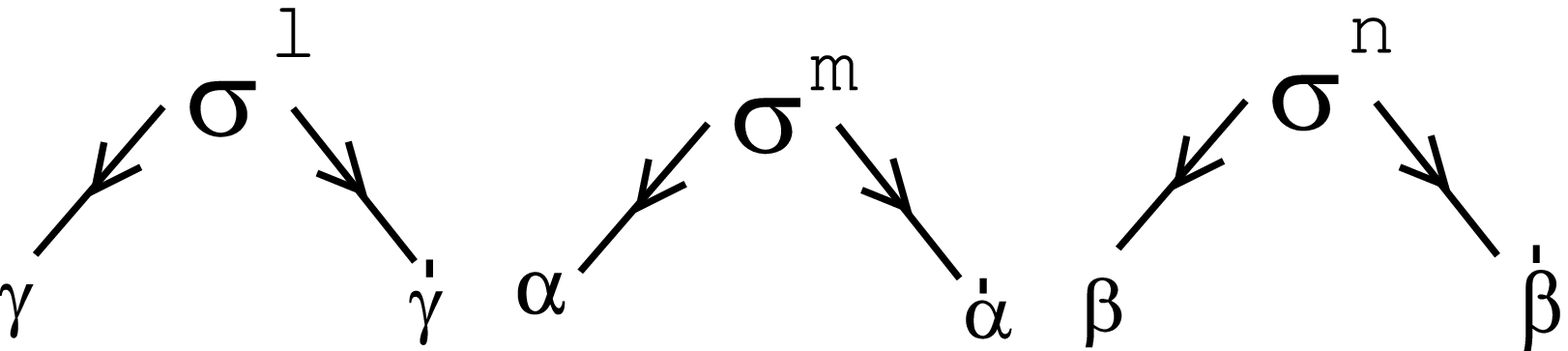}$  \\
\cline{2-4}
  0   &   0      &   1     &  $\graph{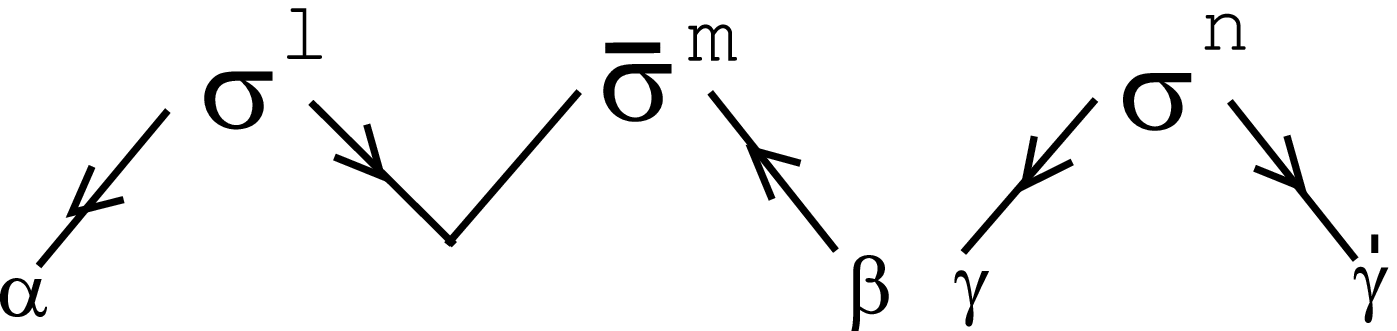}$  \\
\cline{2-4}
       &   1     &   0      &  $\graph{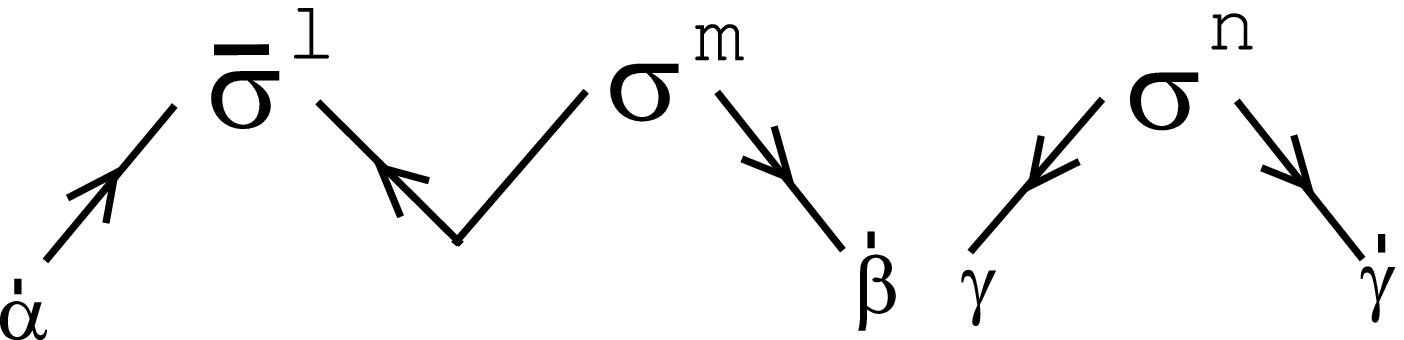}$  \\
\cline{2-4}
       &   1      &   1     &  
\shortstack[l]{
closed-chiral-loop No =1\\
$\graph{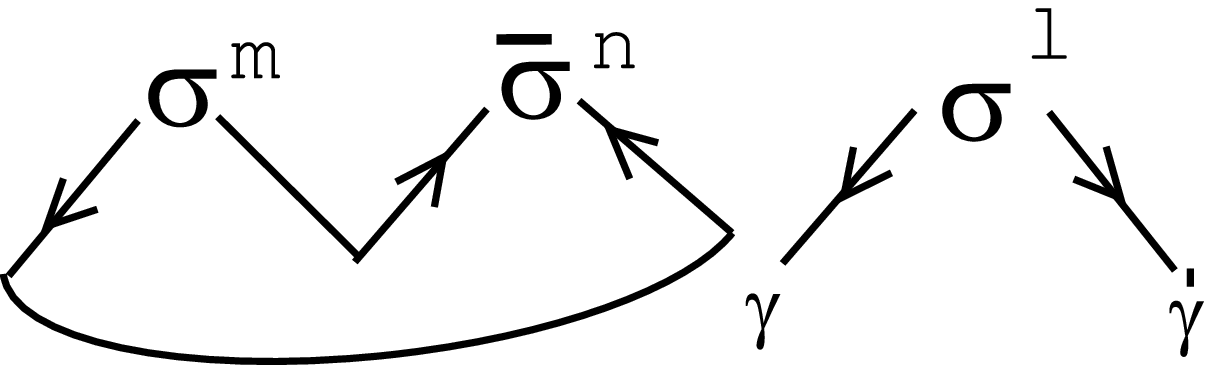}$  }
                                  \\
\cline{4-4}
       &          &         &   
\shortstack[l]{
closed-chiral-loop No =0\\
$\graph{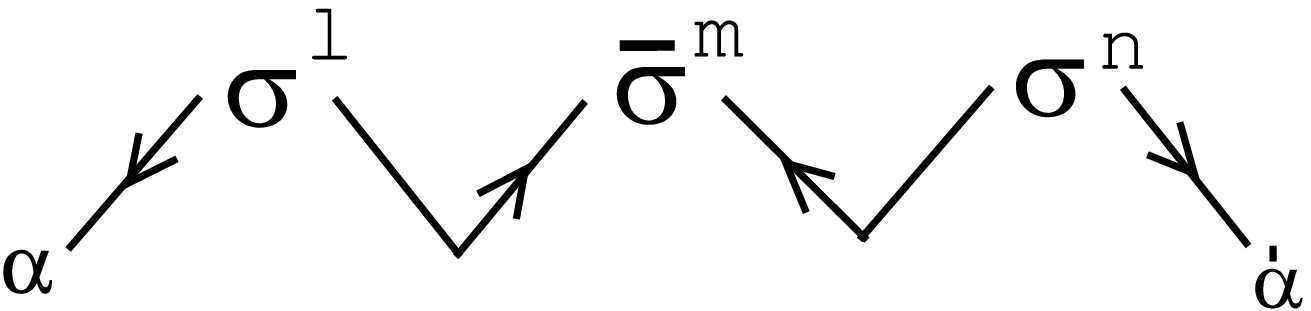}$  }
                                  \\
\hline
       &   0       &  0      &  $-2\ep_{\al\be}\ep_{\aldot\bedot}\ \graph{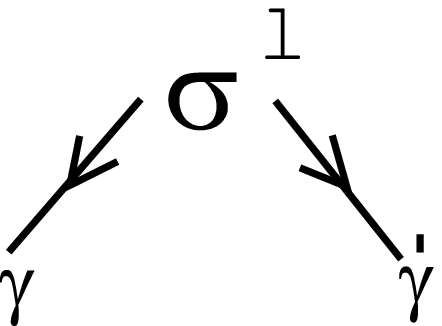}$  \\
\cline{2-4}
   1   &   0      &    1     &  $-4\del^\be_\al\ \graph{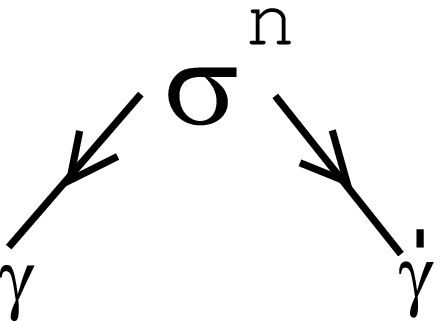}$  \\
\cline{2-4}
       &    1      &    0     &  $-4\del^\aldot_\bedot\ \graph{sig3Dr}$  \\
\cline{2-4}
        &    1     &    1     &   $-8\ \graph{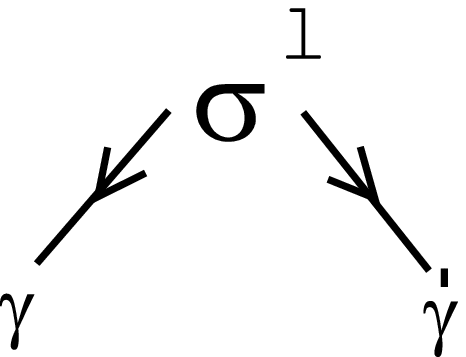}$  \\
\hline
\multicolumn{4}{c}{}\\
\multicolumn{4}{c}{TABLE 2\ Classification of the product of 3 sigma matrices.}
\end{tabular}
\end{center}
\end{table}
These tables clearly show the $\si$-matrices play an important role
to connect the chiral world and the space-time (Lorentz) world. 

Supersymmetry is most manifestly expressed in the superspace 
$(x^m,\sh,\thbar)$. $\sh_\al=\ep_\ab\sh^\be$, 
$\thbar^\aldot=\ep^{\aldot\bedot}\thbar_\bedot$ are spinorial
coordinates. 
They satisfy the relations graphically shown in Fig.4.
\begin{figure}
\resizebox{0.4\textwidth}{!}{%
\includegraphics{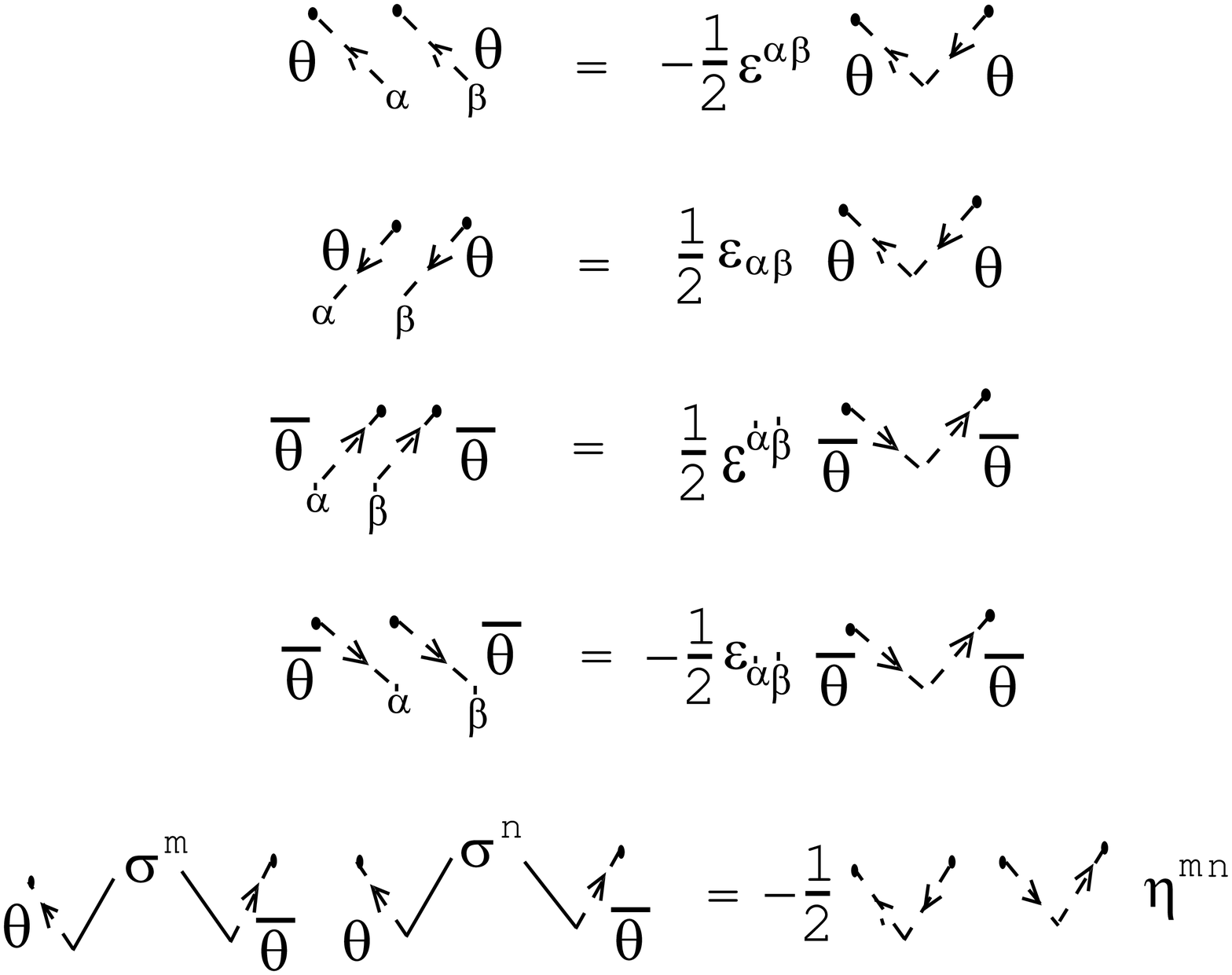}
                             }
\label{LF16}
   \caption{
The graphical rules for the spinor coordinates:\ 
$\sh^\al\sh^\be=-\half\ep^\ab\sh\sh,~\sh_\al\sh_\be=\half\ep_\ab\sh\sh,~
\thbar^\aldot\thbar^\bedot=\half\ep^{\aldot\bedot}\thbar\thbar,~
\thbar_\aldot\thbar_\bedot=-\half\ep_{\aldot\bedot}\thbar\thbar,~
\sh\si^m\thbar\sh\si^n\thbar=-\half\sh\sh\thbar\thbar\eta^{mn}$.
   }
\end{figure}
These relations are exploited in the program in order to 
sort the SUSY quantities with respect to the power of $\sh\sh$ and $\thbar\thbar$. 

For the totally anti-symmetric tensor $\ep^{lmns}$, we introduce
one dimensional array ep[\ ] with 4 components.

\shortstack[l]{
$\mbox{\q\q\q}$\\
$\mbox{\q\q\q}$\\
$\mbox{\q\q\q}$\\
$\mbox{\q\q\q}$
             }  
\q
\shortstack[l]{
ep[0]=l\\
ep[1]=m\\
ep[2]=n\\
ep[3]=s
             }
\q
\shortstack[l]{
$\mbox{\q\q\q}$\\
$\mbox{\q\q\q}$\\
$\mbox{\q\q\q}$\\
$\mbox{\q\q\q}$
             }  
\q
\shortstack[l]{
\\
Symbol: \ e\\
\\
$\mbox{\q\q\q}$
             }
\nl
\nl
This term appears later and produces 
topologically important terms such as
$v_{lm}{\tilde v}^{lm}=\ep^{lmns}v_{lm}v_{ns}$.

As for the metric of the chiral suffix, we do {\it not} introduce specific arrays. They play a role of
raising or lowering suffixes, which can be encoded in the upper (0)
and lower (1) code in arrays. For the Lorentz metric $\eta^{mn}$, we do not
need to much care for the discrimination between the upper and lower suffixes because
of the even-symmetry with respect to the change of the Lorentz suffixes($\eta^{mn}=\eta^{nm}$).

The "reduction" formulae (from the cubic $\si'$s to 
the linear one) are expressed as in Fig.5.
\begin{figure}
\resizebox{0.4\textwidth}{!}{%
\includegraphics{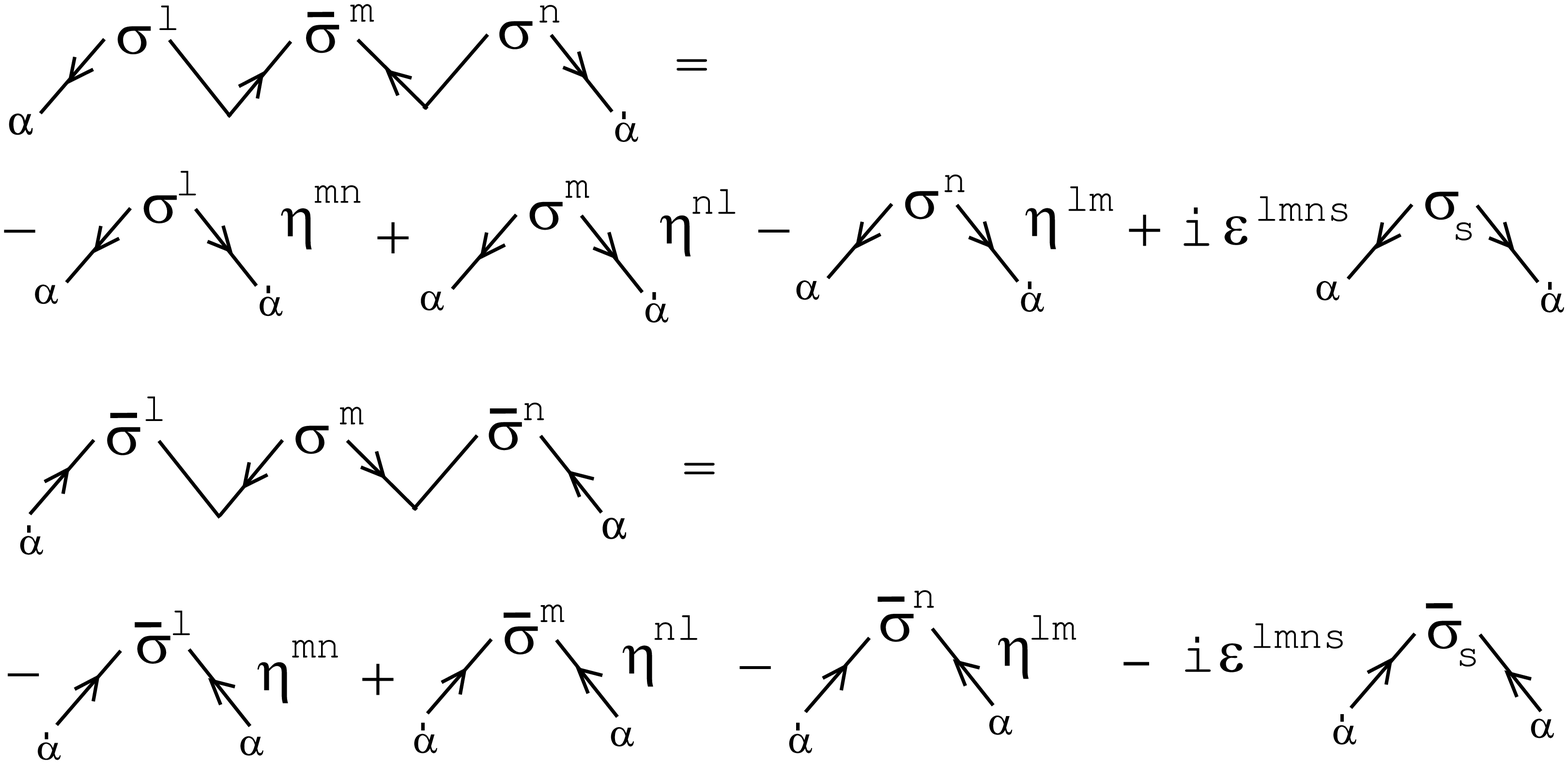}
                             }
\label{LF12R}
   \caption{
Two relations:\ 
1) $\si^l\sibar^m\si^n=-\si^l\eta^{mn}+\si^m\eta^{nl}
-\si^n\eta^{lm}+i\ep^{lmns}\si_s$,\ 
2) $\sibar^l\si^m\sibar^n=-\sibar^l\eta^{mn}+\sibar^m\eta^{nl}
-\sibar^n\eta^{lm}-i\ep^{lmns}\sibar_s$.
   }
\end{figure}
From Fig.5
, we notice any chain of $\si'$s
can always be expressed by less than three $\si'$s. 
The appearance of the 4th rank anti-symmetric tensor $\ep^{lmns}$
is quite illuminating. The following closed chiral-loop graph 
reduces to an interesting quantity.
\begin{eqnarray}
\graphL{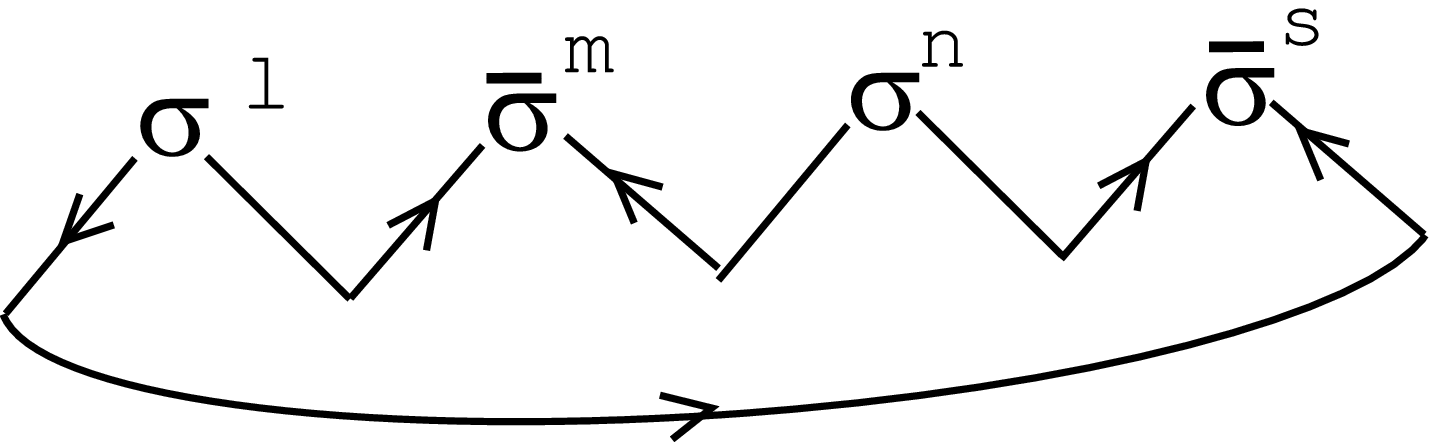}=
2(\eta^{lm}\eta^{ns}-\eta^{ln}\eta^{ms}+\eta^{ls}\eta^{mn}\nn
-i\ep^{lmns})
\label{ssbssbBR2}
\end{eqnarray}

The transformation between
the superfield expression and the fields-components expression is an
important subject of SUSY theories. 
For the
purpose, we do the calculation of $\Phi^\dag\Phi$. In this
case, the input data is taken from the content of the superfield. 
\begin{eqnarray}
\Phi^\dag=-i\graph{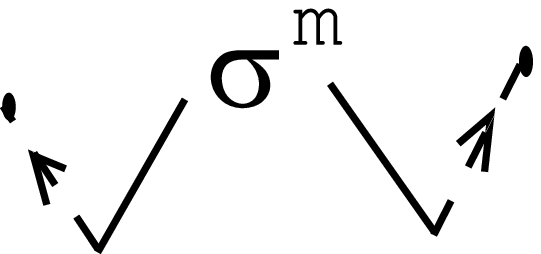}\pl_mA^*+\graph{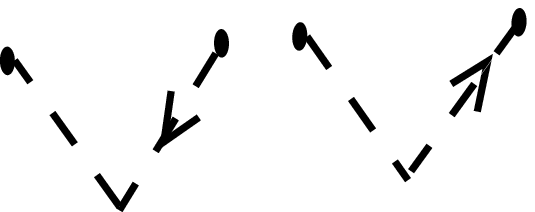}\fourth\pl^2A^*\nn
+2\graph{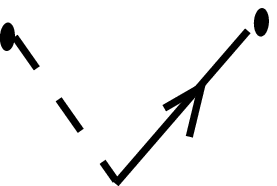}
+i\graph{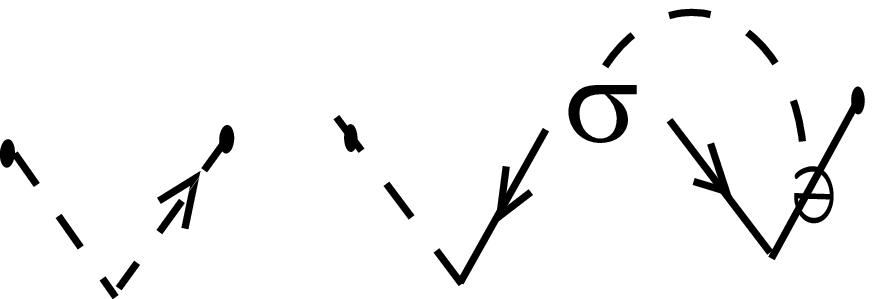}+\graph{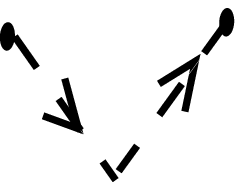}F^*+A^*
\label{PHIB}
\end{eqnarray}
where the directed dotted line is the superspace coordinate $\theta^\alpha$. 
This data of $\Phi^\dag$ is stored as

\shortstack[l]{
weight[sf=0,t=0]=0+i(-1)\\
type[sf=0,t=0,c=0]= t\\
th[sf=0,t=0,c=0,0,0]=1\\
type[sf=0,t=0,c=1]= s\\
si[sf=0,t=0,c=1,0,1]=1\\
si[sf=0,t=0,c=1,1,1]=2\\
siv[sf=0,t=0,c=1]=51\\
type[sf=0,t=0,c=2]= t\\
th[sf=0,t=0,c=2,1,0]=2\\
type[sf=0,t=0,c=3]= B\\
B[sf=0,t=0,c=3,1]=51
             }
\q
\shortstack[l]{
weight[sf=0,t=1]=1+i(0)\\
type[sf=0,t=1,c=0]= t\\
th[sf=0,t=1,c=0,0,0]=1\\
type[sf=0,t=1,c=1]= t\\
th[sf=0,t=1,c=1,0,1]=1\\
type[sf=0,t=1,c=2]= t\\
th[sf=0,t=1,c=2,1,1]=2\\
type[sf=0,t=1,c=3]= t\\
th[sf=0,t=1,c=3,1,0]=2\\
type[sf=0,t=1,c=4]= C\\
C[sf=0,t=1,c=4,1]=1
               }
\q$\cdot\cdot\cdot$\nl
\nl
The calculation of $\Phi\dag\Phi$ leads to the Wess-Zumino Lagrangian. 
\begin{eqnarray}
\Phi^\dag\Phi|_{\sh^2\thbar^2}=-\half \pl_mA^*\pl^mA\ 
+\fourth\pl^2A^*\cdot A\ +\fourth A^*\pl^2A\nn
-i\times(-1)\graph{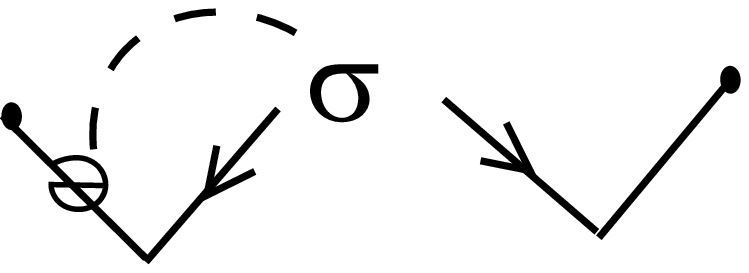}\ +i\times (-1)\graph{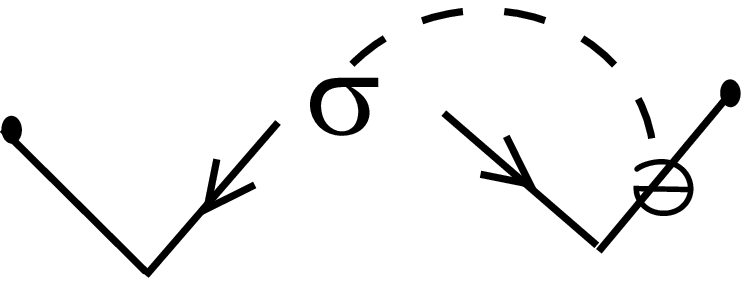}\ 
+F^*F
\label{PHIBPHI}
\end{eqnarray}
We donot ignore the total divergence here.

We also do the calculation of $W_\al W^\al$ where $W_\al$ is the
field strength superfield. They are expressed as follows.\nl
\begin{eqnarray}
W_\al=
-i\graphS{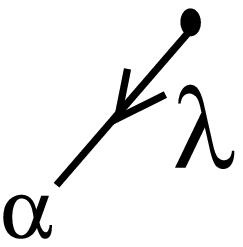}+\graphS{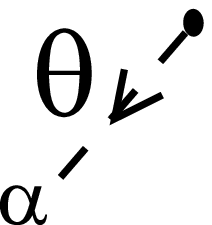}D\nn
-i\graphL{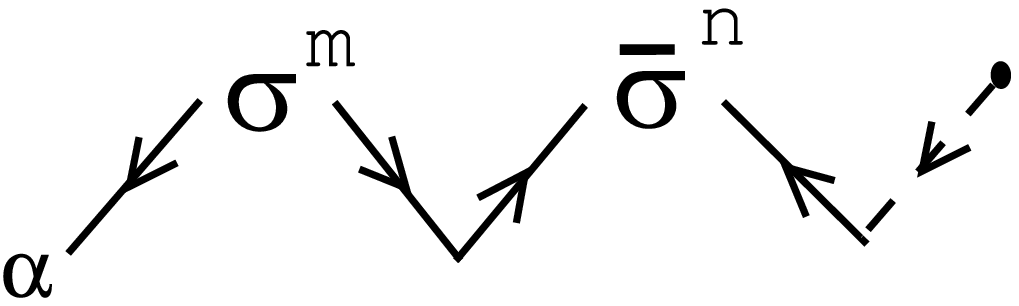}\half v_{mn}+\graphL{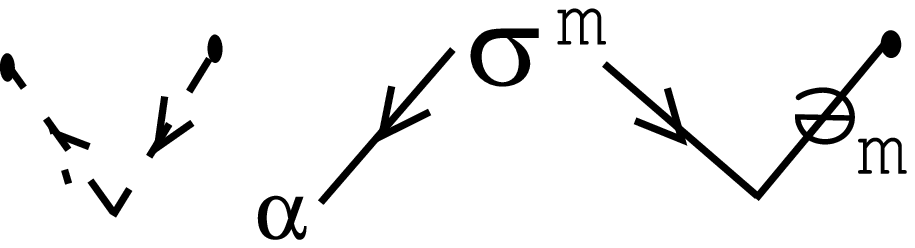}
\label{WalD}
\end{eqnarray}
This data of $W_\al$ is stored as

\shortstack[l]{
weight[sf=0,t=0]=0+i(-1)\\
type[sf=0,t=0,c=0]= l\\
la[sf=0,t=0,c=0,0,1]=1
               }
\q
\shortstack[l]{
weight[sf=0,t=1]=1+i(0)\\
type[sf=0,t=1,c=0]= t\\
th[sf=0,t=1,c=0,0,1]=1\\
type[sf=0,t=1,c=1]= D\\
D[sf=0,t=1,c=1]=1
               }
\q$\cdot\cdot\cdot$\nl

The kinetic term of the photon and the photino, in the SuperQED, is given by
\begin{eqnarray}
\Lcal=\fourth (-W_\al W^\al|_{\sh^2}+\Wbar_\aldot \Wbar^\aldot|_{\thbar^2})\nn
=-\fourth {v_{mn}}^2+\frac{i}{2}(\graph{OutFig2r}-\graph{OutFig1r})+\half D^2\ ,
\label{superEM}
\end{eqnarray}
where we do not ignore the total divergence. 

In the history of the quantum field theory, 
new techniques have produced physically important results. 
The regularization techniques are such examples. 
The dimensional regularization by 'tHooft and Veltman\cite{TV72} produced
important results on the renormalization group property of Yang-Mills theory
and many scattering amplitude calculations. 
The lattice regularization in the gauge theory 
revealed non-perturbative features of hadron physics. 
In this case, the computor technique of numerical calculation
is essential. 
As for the computer algebraic one, we recall the calculation of 2-loop on-shell
counterterms of pure Einstein gravity\cite{GS85,Ven92}. 
A new technique is equally important as a new idea.
  
The SUSY theory is beautifully constructed respecting the symmetry
between bosons and fermions, but the  attractiveness
is practically much reduced by its complicated structure: many fields, 
chiral properties, Grassmannian algebra, etc.  
The present approach intends to improve
the situation by a computer program which makes use of the graphical technique. 
(This approach is taken in Ref.\cite{SI98IJMPC} for the calculation  of product of SO(N) tensors.
It was applied to various anomaly calculations. )

The present program should be much more improved. Here we cite the prospective
final goal.
\begin{enumerate}
\item
It can do the transformation between the superfield expression and
the component expression.
\item
It can do the SUSY trnasformation  
of various quantities. In particular it can confirm the SUSY-invariance 
of the Lagrangian in the graphical way and give the final total divergence.
\item
It can do algebraic SUSY calculation involving $D_\al, \Dbar^\aldot, Q_\al$ and $\Qbar^\aldot$.
\end{enumerate}
The item 1 above has been demostrated in the present paper for the simple
cases of Wess-Zumino model and the Super QED.

It is impossible to deal with all SUSY calculations. This is simply because
which fields appear and which dimensional quantities are calculated
depend on each problem. If we obatin a list of (graph) indices which classify
all physical quantities (operators) appearing in the output, then the present
program works (by adding new lines for the new problem).



\begin{thebibliography}{bib}
\bibitem{SI03}
S. Ichinose, hep-th/0301166, DAMTP-2003-8, US-03-01,
"Graphical Representation of Supersymmetry"
\bibitem{SUSY2004}
S. Ichinose, hep-th/0410027, Proc. 12th Int.Conf. on "Supersymmetry and
Unification of Fundamental Interactions"(June 17-23,2004,Epochal Tsukuba
Congress Center,Japan), p853-856, 
"Graphical Representation of Supersymmetry and Computer Calculation"
\bibitem{SI06UW07}
S. Ichinose, hep-th/0603214, Univ. Vienna preprint UWThPh-2006-7, 
"Graphical Representation of Supersymmetry"
\bibitem{SI06UW08}
S. Ichinose, hep-th/0603220, Univ. Vienna preprint UWThPh-2006-8, 
"Graphical Representation of SUSY and C-Program Calculation"
\bibitem{TV72}
G. 'tHooft and M. Veltman, Nucl.Phys.B44,189(1972)
\bibitem{GS85}
M.H. Goroff and A. Sagnotti, Phys.Lett.B150(1985)81;
Nucl.Phys.B266(1986)709
\bibitem{Ven92}
A.E.M. van de Ven, Nucl.Phys.B378(1992)309
\bibitem{SI98IJMPC} 
S. Ichinose, \IJMP {\bf C9}(1998)243, hep-th/9609014
%
%
%
\end{thebibliography}
\end{document}